\documentclass[screen,sigconf]{acmart}

\AtBeginDocument{%
  \providecommand\BibTeX{{%
    \normalfont B\kern-0.5em{\scshape i\kern-0.25em b}\kern-0.8em\TeX}}}

\copyrightyear{2025}
\acmYear{2025}
\setcopyright{cc}
\setcctype[4.0]{by}

\acmConference[PETRA '25]{The PErvasive Technologies Related to Assistive Environments}{June 25--27, 2025}{Corfu Island, Greece}
\acmBooktitle{The PErvasive Technologies Related to Assistive Environments (PETRA '25), June 25--27, 2025, Corfu Island, Greece}
\acmDOI{10.1145/3733155.3733212}
\acmISBN{979-8-4007-1402-3/25/06}

\usepackage{tabularx}
\usepackage[capitalise,noabbrev]{cleveref}
\usepackage{multirow}

\begin{document}

\title{Identifying the Desired Word Suggestion in Simultaneous Audio}


\author{Dylan Gaines}
\orcid{0000-0002-2747-7680}
\affiliation{%
  \department{Computer Science}
  \institution{Michigan Technological University}
  \city{Houghton}
  \state{Michigan}
  \country{USA}
}
\email{dcgaines@mtu.edu}

\author{Keith Vertanen}
\orcid{0000-0002-7814-2450}
\affiliation{%
  \department{Computer Science}
  \institution{Michigan Technological University}
  \city{Houghton}
  \state{Michigan}
  \country{USA}
}
\email{vertanen@mtu.edu}

\renewcommand{\shortauthors}{Gaines and Vertanen}

\begin{abstract}
We explore a method for presenting word suggestions for non-visual text input using simultaneous voices. We conduct two perceptual studies and investigate the impact of different presentations of voices on a user's ability to detect which voice, if any, spoke their desired word. Our sets of words simulated the word suggestions of a predictive keyboard during real-world text input. We find that when voices are simultaneous, user accuracy decreases significantly with each added word suggestion. However, adding a slight 0.15\,s delay between the start of each subsequent word allows two simultaneous words to be presented with no significant decrease in accuracy compared to presenting two words sequentially (84\% simultaneous versus 86\% sequential). This allows two word suggestions to be presented to the user 32\% faster than sequential playback without decreasing accuracy.
\end{abstract}



\maketitle

\section{Introduction}

Entering text on a device such as a smartphone is a common daily task. Commercial touchscreen keyboards often place a number of word suggestions, commonly three or four, above the keyboard. This allows a user to quickly glance above the keyboard to check the suggestions without unduly distracting from the task of typing. If a user spots their desired word, they can tap it to write it and avoid typing the entire word letter-by-letter. However, for users with visual impairments, it can be a challenge to utilize word suggestions that are presented in such a visual manner. Screen reading software such as Apple's VoiceOver or Google's TalkBack requires a user touch these word suggestions or perform a special gesture to hear them read. This makes the suggestions difficult to discover and can interrupt the natural flow of typing. 

A text entry interface can offer word suggestions when certain conditions are met, such as when the likelihood of a suggestion surpasses a threshold~\cite{montague-inviscid,nicolau-design} or after the user types a certain number of characters~\cite{kristensson-design}. For non-visual interfaces, audio suggestions could be provided automatically when these conditions occur instead of requiring the user to take an action to trigger them. If the user hears the word they are typing they could select it immediately, without needing to finish the remainder of the letters. This would operate similar to many commercially available standard touchscreen keyboards, but using audio instead of visual feedback. 

Reading three or four word suggestions sequentially using text-to-speech (TTS) can take quite some time even with accelerated TTS. The user would likely be able to type the remainder of the word by the time the list was read. To present suggested words more efficiently, we investigate providing the user with audio-only word suggestions using simultaneous voices as suggested by Montague et al.~\cite{montague-inviscid}. This would enable the interface to provide word suggestions much quicker and potentially increase the utility of word predictions for eyes-free text input. 

In this work, we conduct two perceptual studies to measure user performance at detecting a target word among several potentially similar sounding words (as is often the case for a set of word suggestions). We aim to determine not only if users can detect the presence of their target word, but also if they can identify which voice said it as might be required to select their target word in a text entry interface.

We make the following contributions:
\begin{itemize}
    \item We conduct the first quantitative study into whether users can find a target word within a set of similar words played via simultaneous TTS. Our word sets were based on realistic word suggestions made in the context of typing real-world sentences in conjunction with a word prediction algorithm.
    \item We compare the impact of different spatial audio configurations for target word detection in simultaneous TTS audio.
    \item We compare the impact of different start time offsets for target word detection in simultaneous TTS audio.
\end{itemize}

\section{Related Work}

There have been several studies on the human ability to perceive speech amid other speech, the first by Cherry in 1953~\cite{cherry-recognition}. Cherry conducted a series of experiments on the \emph{Cocktail Party Effect}. Cherry found participants could clearly listen to either of two spoken messages recorded by the same person played into opposite ears via headphones. However, participants were not able to notice when the language of the message in the ear that they were not listening to (the ``rejected'' ear) changed from English to German.
This prompted further studies, in which the Cherry found that participants were able to correctly identify if the rejected message was normal human speech (as opposed to reversed speech or an oscillating tone), as well as whether it was a male or female voice. However, they were not able to identify any words in the rejected message or what language it was in. 

This line of work was continued by Egan et al.~\cite{egan-factors} who had participants listen to sentences played simultaneously. Each sentence contained one of two ``call signs''. Participants were instructed to write down the words following a particular call sign, ignoring the sentence that contained the other call sign. In one experiment, they found that high-pass filtering either message allowed participants to better understand the target message. This suggests that users will be able to understand simultaneous speech better if the voices have different sound frequencies. In another experiment, Egan et al. confirmed the results found by Cherry~\cite{cherry-recognition} that participants could understand the target message better if the two messages were played in opposite ears instead of the same ear.  

Brungart~\cite{brungart-masking} conducted experiments using phrases that contained one of eight call signs, followed by one of four colors, and finally one of eight numbers. Two phrases were played simultaneously, and participants were instructed to listen for a target phrase with a particular call sign and report the corresponding color and number. The author varied the speakers of the target phrase and the other masking phrase. They found participants were able to perform better when the speakers were of different sexes. Participants performed the worst when the same speaker was used for both the target and masking phrases, and performance was between these two extremes when the phrases were spoken by different speakers of the same sex. Further work by Darwin et al.~\cite{darwin-fundamental} attribute this performance difference to the combination of differences in vocal tract length and fundamental frequency between speakers.

Brungart and Simpson~\cite{brungart-dynamic} experimented with two, three, and four speakers in different spatial arrangements around the listener. They used digital signal processing and head-related transfer functions to ``move'' the stimuli in space around a user's head. They used the same phrase structure (call sign, color, number) as Brungart~\cite{brungart-masking}. Their results showed that participants were 92\% accurate with two speakers, 72\% with three speakers, and 62\% with four speakers. Additionally, they found that participants scored higher in spatial configurations where the speakers were more spread out, and specifically when the target phrase came from either the left or right side as opposed to the middle.

Guerreiro and Gon\c{c}alves~\cite{guerreiro-text-to-speeches,guerreiro-concurrent} adapted this work to the field of accessible computing, testing how well blind people could understand concurrent speech using screen readers. They used both different-sex voices as well as spatial separation to increase a user's ability to understand the phrases. The authors also increased the speech rate with the aim of presenting as much information as fast as possible while still maintaining comprehension. They found that using two or three voices with slightly faster speech rates had a larger information bandwidth than using a single voice with a much faster speech rate. They reported that the best combination of comprehension and speed was using two voices at either 1.75 or 2 times the default speech rate.

While past studies have explored various factors contributing to the clarity of simultaneous speech, they have all done so using phrases or full sentences. When designing a text entry interface that includes word suggestions, we are more concerned with the clarity of single words. Additionally, perceiving the desired word may be more difficult since word suggestions tend to be more similar to one another than randomly selected words (e.g.~if a user has already typed the prefix of a word, several suggestions will likely start with that prefix). Nicolau et al.~\cite{nicolau-design} investigated concurrency while exploring the design space of nonvisual word completions. In their analysis of users' feedback, Nicolau et al.~noted that word discrimination was the biggest drawback to concurrent word suggestions. However, they only reported qualitative feedback from users and did not report the quantitative accuracy of users with this type of audio. Participant comments from \cite{nicolau-design} suggested that word discrimination of concurrent suggestions might be less of an issue in calm environments, and three of nine participants preferred concurrent suggestions to sequential.

\section{Study 1}

We developed a study to determine how well users were able to comprehend single words among other simultaneous words. We asked participants to wear headphones for the duration of the study. In each trial, we presented users with a target word. They then listened to between two and four voices that each spoke a different word at the same time. The user then selected which voice, if any, they thought spoke the target word. We generated the word audio using the Android operating system's TTS service:
\begin{itemize}
    \item \textsc{Voice 1} --- English US, Male 1
    \item \textsc{Voice 2} --- English US, Female 1
    \item \textsc{Voice 3} --- English US, Male 3
    \item \textsc{Voice 4} --- English US, Female 2
\end{itemize}

There were an equal number of trials in each condition with two, three, and four voices, but the order they occurred within each condition was randomized. There were three counterbalanced conditions in our within-subjects study:
\begin{itemize}
    \item \textbf{\textsc{180Degree}} --- \textsc{Voice 1} and \textsc{Voice 2} played entirely in the participants' left ears, while \textsc{Voice 3} and \textsc{Voice 4} played entirely in the participants' right ears (creating 180 degrees of separation between each pair). 
    \begin{itemize}
        \item For trials with two voices, we used  \textsc{Voice 1} and \textsc{Voice 4}.
        \item For trials with three voices, we used  \textsc{Voice 1}, \textsc{Voice 2}, and \textsc{Voice 4}.
        \item Trials with four voices used all four voices.
    \end{itemize}  
    The aim of this ordering was to maximize both pitch and spatial differences between playing voices. All voices began playing simultaneously.
    \item \textbf{\textsc{60Degree}} --- \textsc{Voice 1} and \textsc{Voice 4} played entirely in the left and right ears, respectively. \textsc{Voice 2} was located 30 degrees to the left of center and \textsc{Voice 3} was 30 degrees to the right of center (creating 60 degrees of spatial separation between each voice). 
    \begin{itemize}
        \item The locations for \textsc{Voice 2} and \textsc{Voice 3} were simulated by splitting the audio into left and right channels, adjusting the volume of each, merging them back together, and then normalizing the result to the same total volume as the other voices. The individual channel volumes were adjusted by factors of 0.67 on the left and 0.33 on the right for \textsc{Voice 2} and by factors of 0.33 on the left and 0.67 on the right for \textsc{Voice 3}.
    \end{itemize}
    The same combinations of voices were used as in \textsc{180Degree} and all voices began playing simultaneously.
    \item \textbf{\textsc{Sequential}} --- All voices were located directly in front of the participant (equal volume in both ears). Voices were added and played in numerical order (e.g.~if there were only two voices, \textsc{Voice 1} and \textsc{Voice 2} were used). Each voice waited for the previous voice to finish completely before beginning to speak.
\end{itemize}

Prior to the study, we created the trials based on phrases taken from the Enron mobile data set~\cite{vertanen-enron}. To create the words spoken for each trial, we chose a random Enron phrase and a random character position in that phrase. We fed the portion of the phrase up to and including that position (including any partial word) into the VelociTap decoder~\cite{vertanen-velocitap}. The n-best word suggestions returned by the decoder (the $n$ words with the highest probability given the context and any prefix) became the words spoken by each voice, where $n$ was the number of voices in that particular trial. 

We did not allow the random position to be the beginning of the sentence since the decoder would just return the most likely words to start a sentence. If there were not enough words returned by the decoder to fill every voice in a trial, a new phrase and position were chosen. This procedure produced word suggestions sets that were representative of what a user might receive during real-world text entry. The partially completed word or next word (if the position fell between words) became the target word. Once the words were selected, which voices spoke them were randomized so that \textsc{Voice 1} was not always the most likely word returned by the decoder. It is important to note that the target word was not always present in the words that were spoken, as would be the case in real-world text entry (this was the case in 22\% of trials).

\subsection{Procedure}

We conducted this study remotely via a web application. This allowed us to ensure participant safety during the COVID-19 pandemic. The application first walked participants through informed consent and a demographic questionnaire. It then provided instructions for the first of the three counterbalanced conditions; the ordering of conditions was determined based on participant number. 

\begin{figure}[tb]
  \begin{center}
  \fbox{\includegraphics[width=8.2cm]{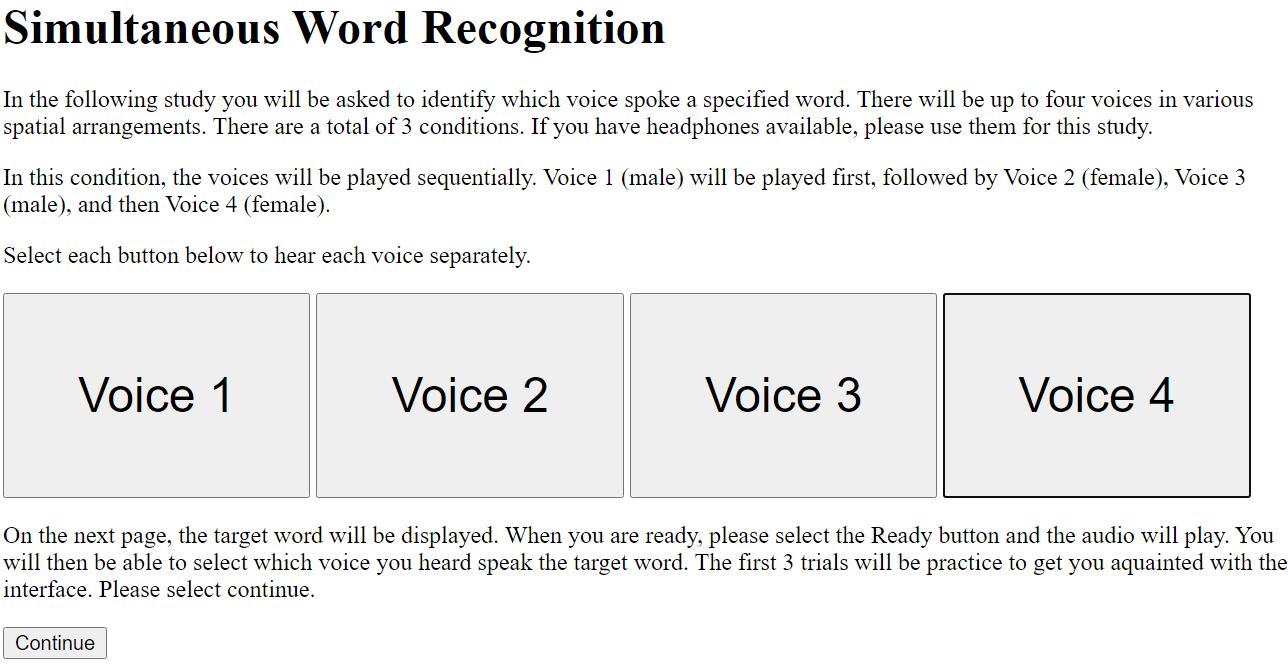}}
  \end{center}
    \vspace{-1mm}
  \caption{The web interface at the beginning of the \textsc{Sequential} condition. The text below the Voice buttons as well as the Continue button did not appear until all four Voice buttons have been clicked. This screen is meant to familiarize users with which voice corresponds to which number.}
  \Description{The instructions read, ``In the following study you will be asked to identify which voice spoke a specified word. There will be up to four voices in various spatial arrangements. There are a total of 3 conditions. If you have headphones available, please use them for this study. In this condition, the voices will be played sequentially. Voice 1 (male) will be played first, followed by Voice 2 (female), Voice 3 (male), and then Voice 4 (female). Select each button below to hear each voice separately.'' Below the button for each voice, the instructions continue, ``On the next page, the target word will be displayed. When you are ready, please select the Ready button and the audio will play. You will then be able to select which voice you heard speak the target word. The first 3 trials will be practice to get you acquainted with the interface. Please select continue.''}
  \label{instructions-figure}
\end{figure}

To familiarize participants with the voices and their locations for each condition, we included a button for each voice on the instruction page (\cref{instructions-figure}). When clicked, these buttons played audio speaking their voice number (e.g.~Voice 1 played the text-to-speech for ``Voice 1''). We synthesized the voice number using the corresponding voice and spatially located each voice in its position for the current condition. We required participants click all four buttons before they could proceed. They were allowed to listen to each voice as many times as they wanted. The instructions also asked participants to wear headphones for the study. Participants completed this instruction process at the start of each condition.

\begin{figure}[tb]
  \begin{center}
  \fbox{\includegraphics[width=8.2cm]{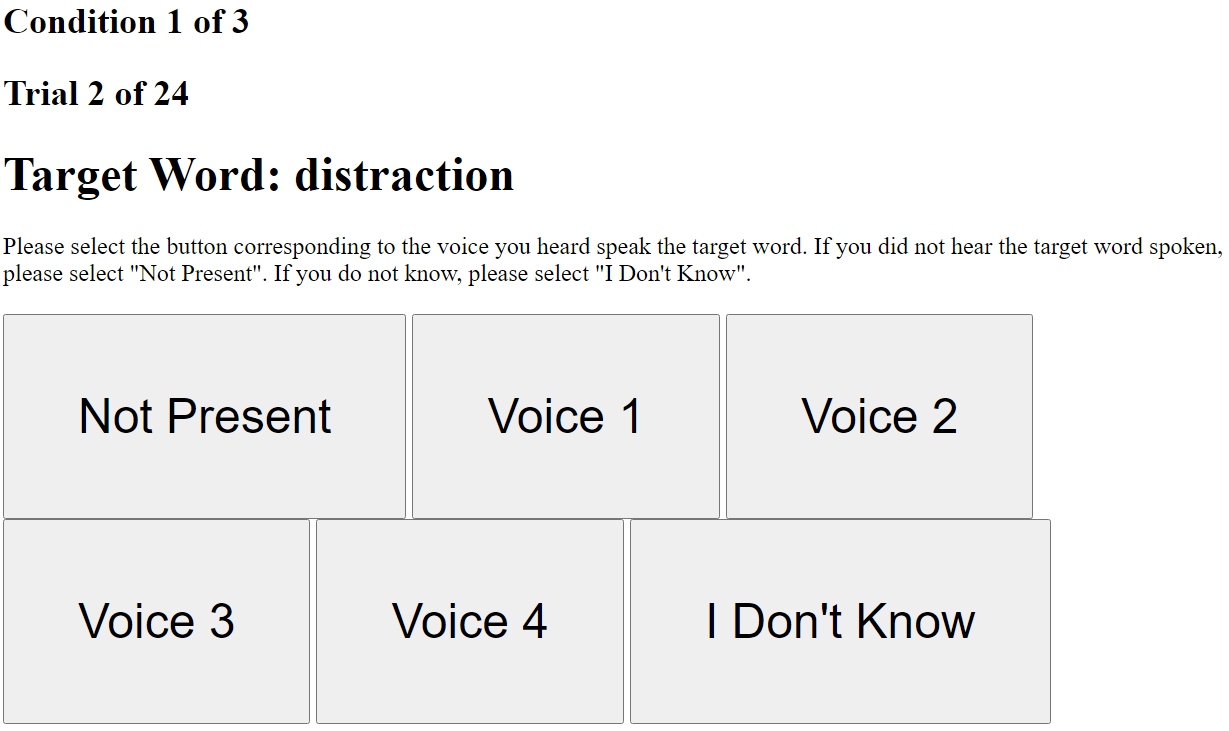}}
  \end{center}
      \vspace{-1mm}
  \caption{The web interface once the participant had clicked `Ready' and the audio had played. The participant now selects which voice they heard say the target word.}
  \Description{The web interface displayed after a participant indicated they were ready and listened to the audio. The interface displays the condition and trial number, and the target word. It contains instructions that read ``Please select the button corresponding to the voice you heard speak the target word. If you did not hear the target word spoken, please select Not Present. If you do not know, please select I Don't Know.'' Below are six buttons reading ``Not Present'', ``Voice 1'', ``Voice 2'', ``Voice 3'', ``Voice 4'', and ``I don't know''.}
  \label{interface-figure}
\end{figure}

Next, participants completed six practice trials. They were first shown the target word. When the participant was ready, they clicked a button and the audio for the trial was played. They were shown six buttons: ``Not Present'', the four voices, and ``I Don't Know'' (see \cref{interface-figure}). To keep the interface consistent between trials, all four voice buttons were shown regardless of the number of voices in the trial's audio. If the participant answered correctly, they proceeded to the next trial. Otherwise, the correct response was shown and the participant repeated the practice trial until they responded correctly. The purpose of this was to ensure that participants were familiar with the voices that were being presented. 

Practice trials were excluded from our analysis. After the practice trials, participants completed 24 evaluation trials. The interface was identical except participants were unable to retry incorrect trials. At the end of each condition, participants completed a brief questionnaire. They then repeated this process for the next condition, beginning with the instructions page.

\begin{figure}[tb]
  \begin{center}
  \includegraphics[width=8.2cm]{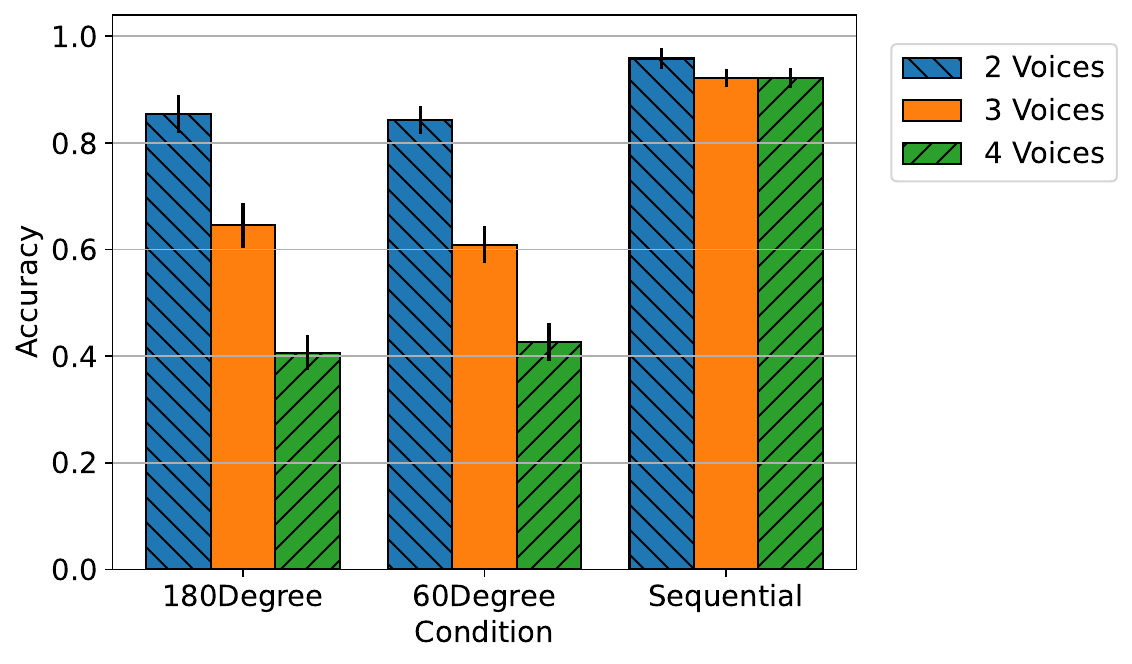}
  \end{center}
  \caption{The average accuracy of participants in Study 1. Error bars represent standard error of the mean.}
  \Description{A bar graph showing the accuracy of participants in each condition. In the 180 Degree condition, participants had an accuracy of 85\%, 65\%, and 41\% with two, three, and four voices. In the 60 Degree condition, they had accuracies of 84\%, 61\%, and 43\%. In the sequential condition, they had accuracies of 96\%, 92\%, and 92\%.}
  \label{study-1-graph}
\end{figure}

\subsection{Results}

We recruited 24 participants by word-of-mouth. A total of 5 participants did not fully complete the study due to technical issues. We removed their data and recruited new participants to obtain a total of 24 complete participants. We paid each participant US\$10 for their participation. Participants ranged from 18--25 years of age. 9 identified as male, 14 as female, and 1 as other. None of the participants reported having uncorrected hearing or visual impairments.
\begin{table*}[tb]
  \begin{center}
    \begin{tabular}{ l c r r r r r r r r r }
    \toprule
           & & \multicolumn{9}{c}{Condition} \\
           \cmidrule(l){3-11}
           & & \multicolumn{3}{c}{\textsc{180Degree}} & \multicolumn{3}{c}{\textsc{60Degree}} & \multicolumn{3}{c}{\textsc{Sequential}} \\
           \cmidrule(lr){3-5}
           \cmidrule(lr){6-8}
           \cmidrule(lr){9-11}
            Condition & Voices & 2 & 3 & 4 & 2 & 3 & 4 & 2 & 3 & 4 \\ 
    \midrule
                       & 2 & - & - & - & - & - & - & - & - & \hspace{4mm}- \\
    \textsc{180Degree} & 3 & \textbf{.002} & - & - & - & - & - & - & - & - \\
                       & 4 & \textbf{\textless.001} & \textbf{\textless.001} & - & - & - & - & - & - & - \\
    \midrule 
                       & 2 & 1.00 & \textbf{.010} & \textbf{\textless.001} & - & - & - & - & - & - \\
    \textsc{60Degree}  & 3 & \textbf{\textless.001} & 1.00 & \textbf{.007} & \textbf{\textless.001} & - & - & - & - & - \\
                       & 4 & \textbf{\textless.001} & \textbf{.010} & 1.00 & \textbf{\textless.001} & \textbf{.024} & - & - & - & - \\
    \midrule 
                       & 2 & .536 & \textbf{\textless.001} & \textbf{\textless.001} & 0.55 & \textbf{\textless.001} & \textbf{\textless.001} & - & - & - \\
    \textsc{Sequential}& 3 & 1.00 & \textbf{\textless.001} & \textbf{\textless.001} & 1.00 & \textbf{\textless.001} & \textbf{\textless.001} & 1.00 & - & - \\
                       & 4 & 1.00 & \textbf{\textless.001} & \textbf{\textless.001} & .786 & \textbf{\textless.001} & \textbf{\textless.001} & 1.00 & 1.00 & - \\    
    \bottomrule
    \end{tabular}
  \end{center}
  \caption{The p-values from pairwise post hoc t-tests in Study 1. Bold values indicate a significant difference ($p < 0.05$).}
  \label{study-1-results-table}
\end{table*}

Our study had two independent variables: spatial arrangement and number of voices. The dependent variable was response accuracy, which we computed as the number of correct responses divided by the number of total responses. If a participant selected ``I don't know'', this was considered an incorrect response. 

The mean response accuracy for each combination of spatial arrangement and number of voices can be seen in \cref{study-1-graph}. For the \textsc{180Degree} condition, participants had an average accuracy of 85\%, 65\%, 41\% with two, three, and four voices, respectively. The results in the \textsc{60Degree} condition were similar, with an accuracy of 84\% for two voices, 61\% for three voices, and 43\% for four voices. As we expected, the \textsc{Sequential} condition did not have the same decrease in accuracy as voices were added. Participants had accuracies of 96\% (two voices), 92\% (three voices), and 92\% (four voices).

Mauchly's sphericity test showed that both main effects met the assumption of sphericity ($W = 0.96, p = 0.66$ for spatial arrangement and $W > 0.99, p = 0.99$ for number of voices). Additionally, the interaction also met the assumption of sphericity ($W = 0.82, p = 0.90$), so we did not need to correct the degrees of freedom. A type III ANOVA for our 3 (spatial arrangement) $\times$ 3 (number of voices) design showed a significant difference between spatial arrangements, $F(2,46) = 83.80$, $p < .001$, as well as number of voices, $F(2,46) = 102.35$, $p < .001$. More importantly, it showed a significant interaction between our independent variables, $F(4,72) = 19.38$, $p < .001$. This indicates that the number of voices had a different effect on participants' accuracy depending on the spatial arrangement used. Namely, increasing the number of voices decreased performance significantly in the simultaneous conditions, but not in the baseline \textsc{Sequential} condition. Thus, we did not summarize our results within each independent variable.

We ran pairwise \emph{post hoc} t-tests with Bonferroni correction to locate any significant differences. The results of these tests are shown in \cref{study-1-results-table}. Notably, within both \textsc{180Degree} and \textsc{60Degree}, the difference between each number of voices was significant. Trials that had three voices had significantly lower accuracy than trials with two voices, and trials with four voices were significantly lower still. However, within \textsc{Sequential}, the number of voices did not significantly impact accuracy. 

Between the two simultaneous conditions, \textsc{180Degree} and \textsc{60Degree}, differences were significant only where the number of voices was not equal. This suggests that there was not a large difference between these two spatial arrangements. 

Finally, the difference between each simultaneous condition with two voices and \textsc{Sequential} with any number of voices was not significant, suggesting that simultaneous audio could be used if there were only two word suggestions. However, we do not recommend using simultaneous audio for more than two words since three and four voices within the simultaneous conditions were significantly worse than \textsc{Sequential} with any number of voices.

\subsection{Discussion}

Our first study showed that performance did not significantly change when there were only two voices. However, participant accuracy was significantly lower in both simultaneous conditions when there were more than two voices. This was true both when there were two voices and in the \textsc{Sequential} control condition. 

We were hoping to have success similar to past work by creating both spatial and pitch differences in our speakers. However, we conjecture that the difficulty for participants arose from the short length and similarity of the stimuli. All of the previous studies we discussed earlier in this work used sentence or phrase stimuli that may have allowed participants more time to adjust to hearing multiple speakers. 

Additionally, many previous studies, like those by Brungart~\cite{brungart-masking} and Brungart and Simpson~\cite{brungart-dynamic}, used a limited set of words in their phrases that were not easily confusable. This was not the case in our study, and that may have also contributed to the poorer performance observed in the simultaneous conditions. We also reduced the likelihood of participants guessing the correct voice by instructing them to select ``I don't know'' if they were unsure, since it is likely that a user entering text would not elect to select a word suggestion if they were not sure which to select. Previous studies (e.g.~\cite{brungart-dynamic, brungart-masking} did not report having an ``I don't know'' option available. 

\section{Study 2}

We wanted to further explore audio word suggestions to see if we could provide more than two suggestions without sacrificing a user's ability to distinguish them. We devised another experiment that introduced a small delay between the start of words in the simultaneous conditions. This still allowed a series of suggestions to be presented to the user more quickly than fully sequential, but with potentially better comprehension than fully simultaneous. A similar technique was used by Nicolau et al.~\cite{nicolau-design}, but they did not report any quantitative comparisons. Nicolau et al. also only used a 0.25\,s delay; they did not experiment with different delay lengths.

We used the same voices and trial creation process as in the Study 1. All conditions used the same spatial arrangement as the Study 1's \textsc{60Degree} condition. We also added voices in the same order as \textsc{60Degree} to maximize spatial separation (\textsc{Voice 1}, \textsc{Voice 4}, \textsc{Voice 2}, then \textsc{Voice 3}). The conditions for Study 2 were:
\begin{itemize}
    \item \textbf{\textsc{DelayShort}} - Voices began playing in numerical order. After each voice began, there was a 0.15\,s delay before the next voice began playing.
    \item \textbf{\textsc{DelayLong}} - Voices began playing in numerical order. After each voice began, there was a 0.25\,s delay before the next voice began playing.
    \item \textbf{\textsc{Sequential}} - Voices played in numerical order. Each voice waited for the previous voice to finish before beginning to play.
\end{itemize}

\subsection{Procedure}

This study used the same web interface as Study 1. Participants completed informed consent, a demographic questionnaire and received instructions before completing six practice trials and 24 evaluation trials in each condition. There was a short questionnaire after each condition. To recruit a larger and more diverse set of participants in this study, we recruited 96 participants by posting Human Intelligence Tasks (HITs) on Amazon Mechanical Turk, a crowdsourcing platform. We paid participants \$3.50 for their time. 

Similar to the crowdsourcing procedure used by Vertanen et al.~\cite{vertanen-impact}, we eliminated workers that failed to meet a control standard. The standard we chose was 50\% accuracy in the \textsc{Sequential} (control) condition. Since there was no overlap between the voices in the \textsc{Sequential} condition, this should have been an easy accuracy to achieve. We returned rejected HITs to the pool until we had 96 participants that met the standard. No participant was able to complete more than one HIT in this study, and, to our knowledge, no participants had previously completed Study 1 since we recruited via word-of-mouth in Study 1 and not via Amazon Mechanical Turk.

\subsection{Results}\label{study-2-results-sec}

The 96 participants in this study ranged from 21 to 69 years of age. 68 identified as male, and the remaining 28 as female. Three participants identified as being low vision. An additional seven reported using a screen reader either on mobile or desktop devices.

\begin{figure}[tb]
  \begin{center}
  \includegraphics[width=8.2cm]{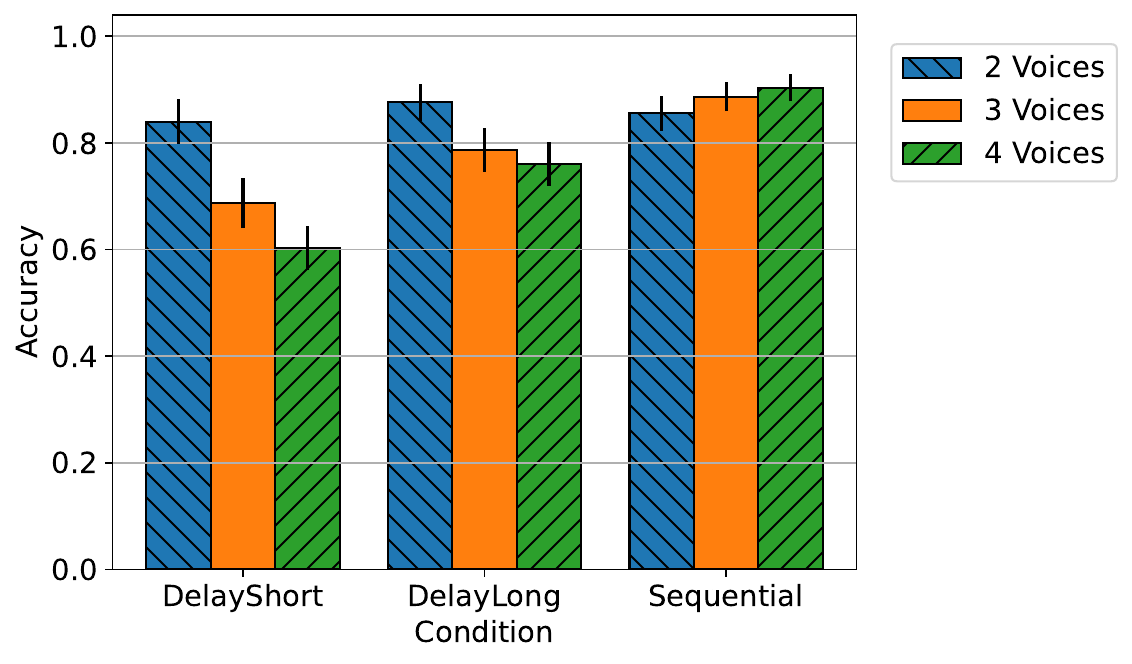}
  \end{center}
  \caption{The average accuracy of participants in Study 2. Error bars represent standard error of the mean.}
  \Description{A bar graph showing the accuracies of participants in each condition of Study 2. In the Delay Short condition, participants had average response accuracies of 84\% with two voices, 69\% with three voices, and 60\% with four voices. In the Delay Long condition, they had accuracies of 88\%, 79\%, and 76\% with two, three, and four voices, respectively. In the \textsc{Sequential} condition, participants had accuracies of 86\% (two voices), 89\% (three voices), and 90\% (four voices).}
  \label{study-2-results-graph}
\end{figure}

\begin{table*}[tb]
  \begin{center}
    \begin{tabular}{ l c r r r r r r r r r }
    \toprule
           & & \multicolumn{9}{c}{Condition} \\
           \cmidrule(l){3-11}
           & & \multicolumn{3}{c}{\textsc{DelayShort}} & \multicolumn{3}{c}{\textsc{DelayLong}} & \multicolumn{3}{c}{\textsc{Sequential}} \\
           \cmidrule(lr){3-5}
           \cmidrule(lr){6-8}
           \cmidrule(lr){9-11}
            Condition & Voices & 2 & 3 & 4 & 2 & 3 & 4 & 2 & 3 & 4 \\ 
    \midrule
                        & 2 & - & - & - & - & - & - & - & - & \hspace{4mm}- \\
    \textsc{DelayShort} & 3 & \textbf{\textless.001} & - & - & - & - & - & - & - & - \\
                        & 4 & \textbf{\textless.001} & \textbf{.042} & - & - & - & - & - & - & - \\
    \midrule 
                        & 2 & 1.00 & \textbf{\textless.001} & \textbf{\textless.001} & - & - & - & - & - & - \\
    \textsc{DelayLong}  & 3 & .228 & \textbf{.002} & \textbf{\textless.001} & \textbf{\textless.001} & - & - & - & - & - \\
                        & 4 & \textbf{.004} & .063 & \textbf{\textless.001} & \textbf{\textless.001} & 1.00 & - & - & - & - \\
    \midrule 
                        & 2 & 1.00 & \textbf{\textless.001} & \textbf{\textless.001} & 1.00 & .063 & \textbf{.004} & - & - & - \\
    \textsc{Sequential} & 3 & 1.00 & \textbf{\textless.001} & \textbf{\textless.001} & 1.00 & \textbf{\textless.001} & \textbf{\textless.001} & 1.00 & - & - \\
                        & 4 & \textbf{.015} & \textbf{\textless.001} & \textbf{\textless.001} & 1.00 & \textbf{\textless.001} & \textbf{\textless.001} & .255 & 1.00 & - \\    
    \bottomrule
    \end{tabular}
  \end{center}
  \caption{The p-values from pairwise post hoc t-tests in the Study 2. Bold values indicate a significant difference ($p < 0.05$).}
  \label{study-2-post-hoc-table}
\end{table*}

In this study, our independent variables were the length of the delay and the number of voices. As with Study 1, our dependent variable was accuracy. The mean accuracy results broken down by each combination of delay and number of voices can be seen in \cref{study-2-results-graph}. In the \textsc{DelayShort} condition, participants had average response accuracies of 84\% with two voices, 69\% with three voices, and 60\% with four voices. Increasing the delay to 0.25\,s resulted in accuracies of 88\%, 79\%, and 76\% with two, three, and four voices, respectively. The baseline \textsc{Sequential} condition yielded accuracies of 86\% (two voices), 89\% (three voices), and 90\% (four voices). 

It was interesting to note that accuracy increased with the number of voices in the \textsc{Sequential} condition. One possible explanation for this is that participants may have selected Voice 2, for example, if the second voice spoke their word, even if only Voices 1 and 4 were speaking, as was the case when only two voices were present. Participants selected a voice that was not a valid option infrequently, with this occurring in only 6.4\% of trials with less than four voices across all conditions. However, we did notice this was the highest in \textsc{Sequential} with two voices at 10.5\% of trials. Further clarification in the instructions or disabling the buttons for voices that were not a part of a particular trial may reduce this occurrence in future work.


As in the previous study, Mauchly's sphericity test showed that both main effects met the assumption of sphericity ($W = 0.99, p = 0.65$ for delay length and $W = 0.99, p = 0.64$ for number of voices). Additionally, the interaction also met the assumption of sphericity, $W = 0.86, p = 0.11$, so once again we did not need to correct the degrees of freedom. A type III ANOVA for our 3 (delay length) $\times$ 3 (number of voices) design showed a significant difference between delay length, $F(2,190) = 78.42$, $p < .001$, and between number of voices, $F(2,190) = 37.60$, $p < .001$. Again, there was a significant interaction between our independent variables, $F(4,380) = 30.13$, $p < .001$, showing that the number of voices had a different impact on accuracy for different delays.

We ran pairwise \emph{post hoc} t-tests with Bonferroni correction to locate any significant differences. These results are presented in \cref{study-2-post-hoc-table}. While, as we noted, mean accuracy increased slightly with number of voices in the \textsc{Sequential} condition, none of these differences were significant. In the \textsc{DelayLong} condition, participants' accuracy with two voices was significantly higher than with three or four, but there was no significant difference between three and four voices. In the \textsc{DelayShort} condition, accuracy significantly decreased with each voice added. 

The most interesting difference from Study 1 was that the detection accuracies of two voices in the \textsc{DelayShort} condition and two or three voices in the \textsc{DelayLong} condition were not statistically significant from two voices in \textsc{Sequential}. While we cannot conclude from this that there was no difference between these conditions, the large number of participants in our study suggests that any difference in accuracy is likely to be small. It may be worth further investigating if the faster playback is worth any accuracy difference that may exist in a text entry task.

To measure the time saved by using simultaneous audio, we averaged the duration of each trial's audio. This can be found, broken down by condition and number of voices, in \cref{study-2-playing-time-table}. As expected, this was the lowest in \textsc{DelayShort}, with an average playing time of 0.95 seconds across all numbers of voices. 
Each trial had one fewer number of delays than it did voices (e.g.~2 delays for a trial with 3 voices), and the difference between the short and long delays was 0.1 seconds. From this, we would expect the \textsc{DelayLong} condition to have a 0.2 second higher average duration than \textsc{DelayShort}. 
We observed just that, with the average playing time of 1.15 seconds for the \textsc{DelayLong} condition. The trials in the \textsc{Sequential} condition had an average playing time of 1.77 seconds, a 54\% increase from the \textsc{DelayLong}.

\begin{table}[tb]
  \begin{center}
    \setlength{\tabcolsep}{4pt}
    \begin{tabular}{c c c c}
    \toprule
           & \multicolumn{3}{c}{Condition} \\
           \cmidrule(l){2-4}
    Voices & \textsc{DelayShort} & \textsc{DelayLong} & \textsc{Sequential} \\
    \midrule
    2 & 0.78 & 0.87 & 1.15 \\
    3 & 0.95 & 1.16 & 1.76 \\
    4 & 1.13 & 1.43 & 2.40 \\
    \midrule
    Average & 0.95 & 1.15 & 1.77 \\
    \bottomrule
    \end{tabular}
  \end{center}
  \caption{The average playing time in seconds for each condition and for different number of simultaneous voices.}
  \label{study-2-playing-time-table}
\end{table}

\subsection{Discussion}

Adding a slight delay before starting to play each subsequent voice produced much more favorable results than the spatial and pitch distance alone, especially for a greater number of voices. While it is difficult to directly compare the \textsc{DelayShort} and \textsc{DelayLong} conditions to the \textsc{60Degree} condition from Study 1 given the different participant pools, the mean accuracies were much higher: 60.2\% and 76.0\% with four voices in \textsc{DelayShort} and \textsc{DelayLong}, respectively, compared to 42.7\% in \textsc{60Degree} from Study 1. 

Most importantly, as noted in \cref{study-2-results-sec}, \textsc{DelayLong} with two or three voices did not have a statistically significant difference in accuracy from \textsc{Sequential} with two voices. This shows that we were able to provide two suggestions 24\% faster (0.87\,s versus 1.15\,s) or an additional suggested word in approximately the same amount of time (1.16\,s versus 1.15\,s) without significantly decreasing detection accuracy. Further, \textsc{DelayShort} with two voices had a similar accuracy of 84\% compared to \textsc{Sequential} with two voices of 86\%. Using a short delay and simultaneous audio was 32\% faster (0.78\,s versus 1.15\,s) than sequential playback.

\subsection{Limitations}
While the results shown in our studies (particularly Study 2) are promising, they are still only a simplification of real-world text entry. Our studies allowed participants to focus all of their attention on listening to the suggestions for the target word. In a more realistic task, users attention will be split between not only the word suggestions, but also the content that they are writing and the act of performing the input itself. This will increase the cognitive load, and we suspect this may further decrease user performance. 

An additional limitation is that while we instructed participants to wear headphones, we do not know for sure if participants followed this instruction. Additionally, in a real-world task, wearing headphones may not always be practical, which could impair users' ability to hear separate left and right audio channels. 

Our study consisted of a single study session lasting less than an hour. The ability to successfully detect a target word in simultaneous TTS audio may require more time to learn. It would be interesting to conduct a longitudinal study to investigate how user performance changes with practice. We also did not specifically recruit blind or low vision participants, who likely have more experience listening to TTS audio. Additional follow-up work is required to determine how well these results generalize to blind and low vision users.

\section{Conclusion}

We explored different ways to present word suggestions to users using only TTS audio feedback. To our knowledge, we provide the first quantitative results showing if users can perform well in the challenging task of detecting a single target word within a set of similar sounding words. In our first study, we found playing two words simultaneously using spatial audio had a lower detection accuracy of 84--85\% compared to 96\% for sequential playback (though this difference was not significant). Playing three or four words at the same time was significantly worse than sequential playback. 

In our second study, we compared adding small delays before starting each simultaneous voice. We found that by adding a short 0.15\,s delay, users detected the target word about equally as well at 84\% using two simultaneous voices versus 86\% using two fully sequential voices. Playing two words simultaneously made audio feedback 32\% faster. Based on our two studies, if simultaneous audio is to be used for short audio feedback events, we suggest using spatial audio, a small start time offset, and to limit the number of voices to two. 


\begin{acks}
This work was supported by NSF IIS-1909248 and by NSF Graduate Research Fellowship DGE-2034833. We thank Ricardo Gonzalez for pilot testing the web application and suggesting the use of a delay in the speech.
\end{acks}

\balance

\bibliographystyle{ACM-Reference-Format}
\bibliography{references}

\end{document}